# Electronic tuneability of a structurally rigid surface intermetallic and Kondo lattice: CePt$_5$ / Pt(111)

C. Praetorius,[1] M. Zinner,[1] A. Köhl,[1] H. Kießling,[1] S. Brück,[1] B. Muenzing,[1] M. Kamp,[2,3]
T. Kachel,[4] F. Choueikani,[5] P. Ohresser,[5] F. Wilhelm,[6] A. Rogalev,[6] and K. Fauth[1,3,*]

[1]*Physikalisches Institut, Universität Würzburg, Am Hubland, 97074 Würzburg, Germany*
[2]*Technische Physik, Physikalisches Institut, Universität Würzburg, Am Hubland, 97074 Würzburg*
[3]*Wilhelm Conrad Röntgen-Center for Complex Material Systems (RCCM),*
*Universität Würzburg, Am Hubland, 97074 Würzburg, Germany*
[4]*Institut für Methoden und Instrumentierung der Forschung mit Synchrotronstrahlung,*
*Helmholtz-Zentrum Berlin für Materialien und Energie GmbH,*
*Albert-Einstein-Strasse 15, 12489 Berlin, Germany*
[5]*Synchrotron SOLEIL, L'Orme des Merisiers, Saint-Aubin-BP48, 91192 Gif-sur-Yvette CEDEX, France*
[6]*The European Synchrotron, 71 avenue des Martyrs, Grenoble, 38000 France*
(Dated: October 26, 2018)

We present an extensive study of structure, composition, electronic and magnetic properties of
Ce–Pt surface intermetallic phases on Pt(111) as a function of their thickness. The sequence of
structural phases appearing in low energy electron diffraction (LEED) may invariably be attributed
to a single underlying intermetallic atomic lattice. Findings from both microscopic and spectroscopic
methods, respectively, prove compatible with CePt$_5$ formation when their characteristic probing
depth is adequately taken into account. The intermetallic film thickness serves as an effective
tuning parameter which brings about characteristic variations of the Cerium valence and related
properties. Soft x-ray absorption (XAS) and magnetic circular dichroism (XMCD) prove well suited
to trace the changing Ce valence and to assess relevant aspects of Kondo physics in the CePt$_5$
surface intermetallic. We find characteristic Kondo scales of the order of $10^2$ K and evidence for
considerable magnetic Kondo screening of the local Ce $4f$ moments. CePt$_5$/Pt(111) and related
systems therefore appear to be promising candidates for further studies of low-dimensional Kondo
lattices at surfaces.

## I. INTRODUCTION

Atomic impurities in metals retain their magnetic character whenever strong Coulomb repulsion effectively suppresses the double occupancy of the impurity orbital. While the local Coulomb energy is large, the finite interaction of local and itinerant degrees of freedom entails the emergence of small characteristic energy scales[1], leading to unconventional macroscopic behavior at low temperature and complex phase diagrams with competing interactions and orders[2–5]. Typical experimental probes of Kondo and heavy fermion systems such as calorimetry, transport or magnetometry are macroscopic in character and cannot, therefore, directly provide microscopic details of mechanisms and interactions in real or reciprocal space. In this respect, the well-developed machinery of surface science has already proven great potential and provided novel insight in recent years[6–13]. Due to both the intrinsic surface specificity of these methods and the frequent occurrence of modified interactions at surfaces it is oftentimes non-trivial, though, to connect these results to bulk properties of the respective materials[9,10,14–17]. Low-dimensional intermetallic phases at surfaces bear the promising potential of reduced ambiguities in this respect.

Among these, Ce–Pt surface phases on Pt(111) have been repeatedly investigated since the inital report by Tang et al.[18], not only in the attempt to explore correlation physics but also out of interest in their catalytic properties[19–27]. The bulk binary Ce–Pt phase diagram[28,29] exhibits various stable intermetallic phases with ground states ranging from a ferromagnetic Kondo lattice with signs of quantum criticality in CePt$_5$[30] to low temperature antiferromagnetism with unscreened Ce moments in CePt$_5$ according to Ref. 31.

The various studies of the Ce–Pt surface alloys revealed a seemingly rich variety of atomic structures and superstructures which appear as the amount of Ce in the alloying procedure is varied, but also a number of conflicting results[18,20,22,25], as discussed in detail in Ref. 27. Most of the resultant surfaces share a remarkable chemical inertness[18,20,25,32]. This finding, along with hexagonal symmetry and the lattice parameters detected by low energy electron diffraction (LEED) and scanning tunneling microscopy (STM) fostered the assumption that these surfaces share the CePt$_5$ atomic lattice and exhibit a pure Pt termination[20,24,25], despite conflicting results from compositional analysis by x-ray photoelectron spectroscopy (XPS)[18,20] and a reported preference of CePt$_2$ over CePt$_5$ in terms of formation enthalpy[33]. A most recent theoretical study indeed finds a Pt termination to be the lowest energy configuration[34].

The interpretation of findings concerning the electronic structure of the Ce–Pt surface alloys has undergone major reassessments since the initial angle resolved pho-toelectron spectroscopy (ARPES) study by Andrews et al.[19] Lead by the results of their preceding compositial analysis which yielded stoichiometries in the range of $CePt_{2.23...3}$[18], these authors concluded that an intensity enhancement near the Fermi energy $E_F$ observed at $T \approx 120$ K could not possibly derive from a Kondo resonance and seeked alternative explanations. A systematic dependence of the Ce valence on Ce coverage prior to alloying was observed by XPS but not given detailed consideration. Later ARPES work concentrated on a specific intermetallic phase (labeled as structure C′ in Ref. 27 and below). On the basis of better resolved datasets, Garnier et al. – while not making claims with respect to the underlying alloy composition and atomic structure – unanimously identified a low temperature intensity enhancement near $E_F$ as the tail of a Kondo resonance[21,22]. Finally, yet improved experimental conditions allowed Klein et al.[26] to observe incipient coherent heavy band formation in the same phase.

The simultaneous assignment of the $CePt_5$ structure and observation of Kondo physics may come as a surprise given the fact that bulk $CePt_5$ does not appear to display signatures of Kondo physics[35]. If correct, this represents an excellent example of how much bulk and surface physics may actually differ in a given material. At the same time, the experimental situation demonstrates the paramount importance of reliable structural information for achieving a coherent understanding of electronic properties.

In a recent piece of work[27], the structural properties of the Ce–Pt surface intermetallics were comprehensively re-assessed on the basis of extensive STM and LEED experiments. There, it was shown that the different structural phases occurring as a function of initial Ce coverage may indeed be understood as to arise from the formation of $CePt_5$, the seemingly different phases corresponding to well-defined numbers of $CePt_5$ unit cells having formed.

In the present paper, we provide further evidence for the presence of a single characteristic atomic lattice across the range of intermetallic thicknesses studied and demonstrate that both microscopic and spectroscopic results, respectively, are compatible with the stoichiometry of $CePt_5$. On this basis we proceed to discussing electronic and magnetic properties of the $CePt_5$/Pt(111) system as a function of intermetallic thickness. Experimental data were obtained by x-ray absorption (XAS) and circular magnetic dichroism (XMCD) which constitute surface sensitive as well as element and orbital specific probes of electronic structure and local magnetic moments. These characteristics allow us assessing Ce valence, Kondo scale and magnetic Kondo screening within one single set of experiments for a given specimen. In conjunction with the antecedent structural analysis, variations in electronic and magnetic properties are now understood as to arise within a single structural $CePt_5$ phase, where the presence of strain at the interface to the Pt(111) substrate leads to the intermetallic thickness acting as an effective, nonthermal tuning parameter.

## II. EXPERIMENTAL METHODS

The experimental results reported below were obtained on samples prepared in different ultrahigh vacuum (UHV) systems, each of them being equipped with the required facilities for in situ surface preparation, Ce deposition and monitoring and a low energy electron diffraction (LEED) unit. A portable apparatus, designed for magnetic studies involving synchrotron radiation, was used for experimentation at BESSY II and the ESRF, respectively. In auxiliary experiments at SOLEIL the stationary equipment included with the end station was used for sample preparation. Finally, a home lab setup, equipped with a cylindrical mirror analyser (CMA), was used for Auger electron spectroscopy (AES). All units provide base pressures in the low $10^{-10}$ mbar range. During sample preparation, specimen temperatures were determined with an infrared pyrometer.

Bare Pt(111) surfaces were prepared by repeated cycles of 1 keV $Ar^+$ ion sputtering and annealing to 1170 K. Cerium (purity 99.9%, MaTeck GmbH) was evaporated from a thoroughly outgassed tungsten crucible with a commercial electron beam evaporator (Focus EFM3s). All UHV systems feature a quartz microbalance (QMB) which is readily moved into the respective sample positions to obtain a reproducible calibration of the Ce deposition. The Pt(111) substrate was held near ambient temperature during Ce deposition and subsequently annealed to $T \approx 970$ K for 5...10 min. This procedure yields the known, well-ordered and remarkably inert surface reconstructions[18,20,22,25–27].

Over 100 individual preparations were carried out and analyzed. They provide the experimental basis for our understanding of the evolution of the structure of the Ce-Pt surface intermetallic as a function of Ce coverage[27]. Only very few of them resulted in surface reconstructions with azimuthal orientations differing from rotational alignment with the substrate or a rotation by 30°, as observed in some earlier reports[20,24]. We have therefore excluded such specimens from further consideration. As in our previous work[27] we shall denote our specimens by reporting their nominal thickness $t_{nom}$ in multiples of $CePt_5$ unit cells along the surface normal. Geometrically, the Ce coverage required to produce $CePt_5$ amounts to $\approx 0.5$ monolayers per unit cell[20].

$CePt_5$ crystallizes in the hexagonal $CaCu_5$ structure (space group P6/mmm6)[28]. Fig. 1 illustrates the Ce and Pt atomic positions within the $CePt_5$ unit cell as well as the local environment of the Ce sites. Each unit cell extends over two atomic layers along the hexagonal symmetry axis, which coincides with the Pt(111) surface normal in our specimens. The in-plane lattice parameter of bulk $CePt_5$ amounts to $a_{CePt_5} = 5.37$ Å[29], which lies in between the values of $\sqrt{3}d_{Pt}$ and $2d_{Pt}$ of the next neighbor distance $d_{Pt} = 2.77$ Å in the Pt(111) surface. It more closely matches the latter value, though, and a $(2 \times 2)$ superstructure is indeed observed at small initial Ce coverage[25,27]. Each atomic layer in $CePt_5$ contains three atoms per $(2 \times 2)$ unit. Layers with CePt$_2$ composition alternate with pure Pt layers forming a kagome net with the holes arranged such as to co-accommodate the Ce atoms of the adjacent layers.

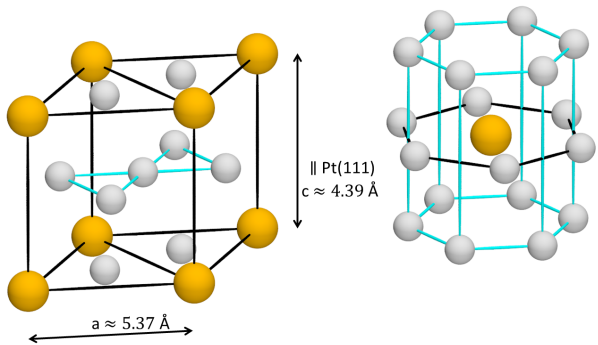

FIG. 1: Unit cell (left panel) and local environment of the Ce atoms (right panel) of bulk CePt$_5$ (CaCu$_5$ structure). Structural parameters of the bulk phase are taken from Ref. 29.

The formation of commensurate superstructures of CePt$_5$/Pt(111) cannot occur without building up strain at the interface and the system indeed forms long ranged superstructures already at small coverage. These are generally interpreted as moiré patterns due to the persistence of a lattice mismatch[20,24,27]. Nevertheless, the atomic density and even more so the density of Pt sites is considerably reduced in CePt$_5$ in comparison to the Pt(111) substrate, a notion to which we shall return in some of our evaluations below.

Each specimen was inspected by LEED at various energies before proceding to further experimentation. Diffractograms indicative of spurious oxidation[24,25,36] or otherwise displaying irreproducible features lead to the elimination of specimens from further analysis. While care was taken in positioning the specimen surface with respect to the LEED optics, this does not warrant the reproducability necessary to reliably assign lattice parameter changes well in the sub-5% range as they occur in the Ce-Pt system. A more precise cross-calibration between the structural phases was achieved by dosing the Ce coverages to the cross-over regimes between adjacent structural phases, as described in detail in Ref. 27.

LEED-IV datasets were produced from series of diffraction patterns recorded in electron kinetic energy increments of 1 eV in the range of $35 \text{ eV} \leq E_{kin} \leq 120 \text{ eV}$ with a consumer camera. Spot intensities were extracted semi-automatically from the series of images using custom developed codes for spot tracking and background subtraction[37].

To obtain morphological and structural information complementary to the UHV based surface science techniques, we have produced a specimen suited for cross sectional scanning transmission electron microscopy (STEM). A film with $t_{nom} \approx 10$ u.c. was transferred to a focussed ion beam (FIB) instrument (Helios Nanolab dual beam, FEI). After deposition of an amorphous pro-tective Pt/C layer a rectangular lamella was cut from the crystal with the 30 keV Ga ion beam, its long side being aligned with a $\langle 1\,1\,\bar{2} \rangle$ direction of the substrate. The lamella was further thinned at progressively decreasing Ga ion energies and subsequently transferred to a dedicated (S)TEM instrument (Titan 80-300 (FEI), resolution 0.136 nm). Scanning transmission electron miccrographs (STEM), viewing along the substrate's $\langle 1\,\bar{1}\,0 \rangle$, were recorded with a high angle annular dark field (HAADF) detector yielding enhanced scattering intensity at higher atomic number.

AES data were produced under equivalent conditions from pristine Pt(111), Ce/Pt(111) deposits before alloying and reconstructed surfaces in normal electron incidence geometry. These measurements targeting stoichiometry determination rather than detailed spectroscopic information, emphasis was put on rapid data acquisition rather than high spectral resolution. For stoichiometry analysis, the non-overlapping but energetically fairly equivalent Auger features at 98/116 eV (Ce) and 158/168 eV (Pt) were chosen.

Soft x-ray Ce M$_{4,5}$ XAS and XMCD measurements were carried out with circular polarized synchrotron radiation during experimental runs at the PM 3 bending magnet beamline of BESSY II[38,39] with a custom set-up ($\pm 3$ T superconducting UHV magnet) and the DEIMOS beamline (local Cryomag $\pm 6.5$ T (7 T) end station)[40] of SOLEIL. Absorption spectra were acquired in the total electron yield (TEY) mode by measuring the sample drain current. Appropriate normalization is provided by simultaneous measurement of the TEY from a gold mesh further upstream.

Using circular polarized x-rays, spectra taken at the 'magic' angle of x-ray incidence ($\theta_X = 54.7°$ off the hexagonal axis) probe the polarization-averaged, 'isotropic' spectrum. This condition is nearly fulfilled in the datasets used for stoichiometry determination ($\theta_X = 60°$), reducing intensity misevaluations due to linear dichroism to a negligible level. The magnetic data presented in this work were taken at normal photon incidence, however. In this geometry the direction of the applied magnetic field coincides with the hexagonal axis. Van Vleck contributions to the magnetic susceptibility therefore vanish[35], rendering magnetic evaluations most straightforward.

Auxiliary hard x-ray absorption datasets were acquired at the Ce L$_{2,3}$ edges on beamline ID12 at the ESRF. In situ sample transfer not having been available there, a specimen of 9 u.c. nominal thickness was prepared and capped with an additional Pt film of 4 ML thickness at ambient temperature. This capping proved sufficient to protect the film from oxidation during ex situ sample transfer. The sample was attached to a cold finger of a constant flow He cryostat and inserted in an experimental chamber. High quality grazing incidence ($\theta_X = 75°$) Ce L$_{2,3}$ edge XANES datasets could be obtained despite the low specimen thickness and Ce concentration by recording the partial fluorescence yield covering L$_{\alpha_{1,2}}$ and L$_{\beta_1}$ emission, respectively, with a 35-channel silicon drift diode detector, developed at the ID12 beamline in collaboration with Eurisys-Mesures (now Canberra Eurisys)[41].

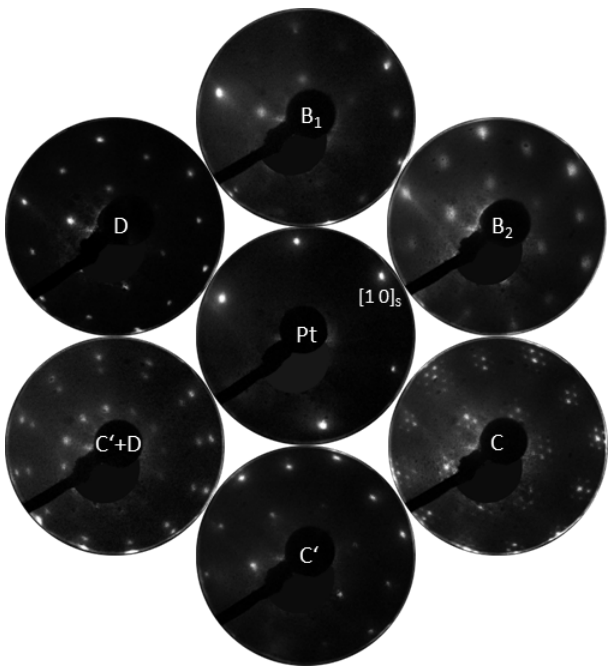

FIG. 2: Compilation of LEED patterns obtained at an electron energy of $E_{kin} = 63$ eV. Pristine Pt(111) (center panel). $(2 \times 2)$ reconstruction at $t_{nom} = 1$ u.c. (B$_1$) and 2 u.c. (B$_2$). $(\frac{10}{9}\sqrt{3} \times \frac{10}{9}\sqrt{3})$R30° with superstructure satellites (C) at 3 u.c. and without satellites at 4...5 u.c. (C′). Superposition of rotated and backrotated (C′+D), compressed $(2 \times 2)$ at 5...10 u.c. and backrotated $(2 \times 2)$ structures only (D) above 10 u.c. thickness.

## III. RESULTS AND DISCUSSION

### A. structure and composition

In a recent study[27] we demonstrated that the essential observations by STM and LEED are well accounted for by assuming CePt$_5$ formation. Nevertheless, the experimental information on constitution and composition remains indirect. Following a brief review of the structural phases observed upon varying the initial Ce coverage, we shall present complementary evidence for rather well defined film formation and discuss the alloy composition on the basis of spectroscopic data. In designating the various surface structures, we adhere to the nomenclature of our previous work[27].

Fig. 2 displays LEED patterns representative of six CePt$_5$ 'phases' we can discern. At small initial Ce coverage ($t_{nom}$ of 1 and 2 u.c. of CePt$_5$, respectively) a $(2 \times 2)$ superstructure forms, aligned with the substrate orientation. The $(2 \times 2)$ spots are significantly broadened, which can be accounted for by the observation of large periodic-ity superstructures in STM (structure B in Refs. 20,27). The corresponding reciprocal lattice remains unresolved in LEED and the surface corrugation gives rise to a distribution of scattering intensity around the $(2 \times 2)$ reciprocal lattice vectors[27].

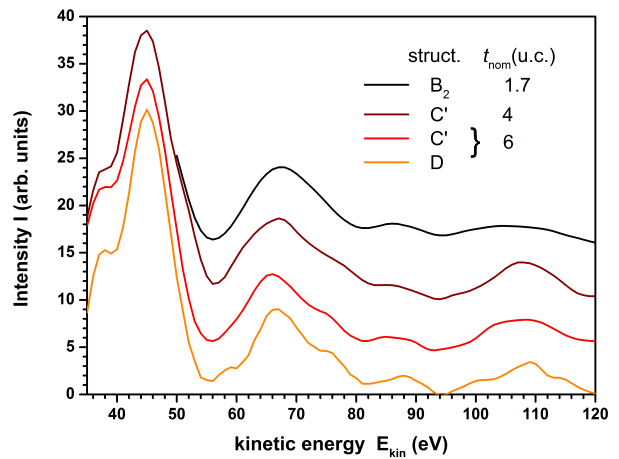

FIG. 3: Averaged energy dependent intensity profiles of the first order "$(2 \times 2)$" diffraction spots collected on structures B$_2$, C′ and D. For the sake of clarity, the different curves have been vertically offset. Their obvious similarity indicates a corresponding similarity of the respective atomic lattices.

The presence of tensile lateral strain can be assessed with LEED patterns taken at higher electron energies, where narrow substrate diffraction peaks are superimposed onto those of CePt$_5$. While the lattice mismatch at $t_{nom} = 1$ u.c. is too small to be quantified, a relaxation of $\approx 1.5\%$ is perceptible at $t_{nom} = 2$ u.c.

Further increasing the Ce coverage leads to the formation of rotated CePt$_5$ films, going along with a marked reduction of the lattice parameter (C). A superstructure again forms, exhibiting a $(\frac{10}{9}\sqrt{3} \times \frac{10}{9}\sqrt{3})$R30° registry between the substrate and CePt$_5$ lattices[27]. Its corrugation is markedly reduced at $t_{nom} = 4$ u.c., resulting in strongly reduced intensities of the LEED satellites (C′).

Upon increasing the CePt$_5$ thickness to above 5 u.c., a second phase with the same lattice parameter appears, again in rotational alignment with the substrate (D). Its diffraction peaks progressively grow in intensity at the expense of those of structure C′ until eventually, for $t_{nom} > 10$ u.c., it is the only structure observed. This remains true up to at least 15 u.c., which is the upper limit of the thickness range studied.

More information about the atomic structure can be obtained from energy dependent LEED intensities. A full LEED-IV analysis is beyond the scope of the present paper and shall be presented elsewhere[42]. It does provide evidence, however, for a Pt terminated CePt$_5$ film at $t_{nom} = 4$ u.c. with a lattice parameter of 5.35 Å and additional Pt atoms residing in the kagome hole positions. The latter finding is in perfect agreement with a recent theoretical analysis of the surface energetics[34] and offers a natural explanation for the chemical inertness of the alloy surfaces.

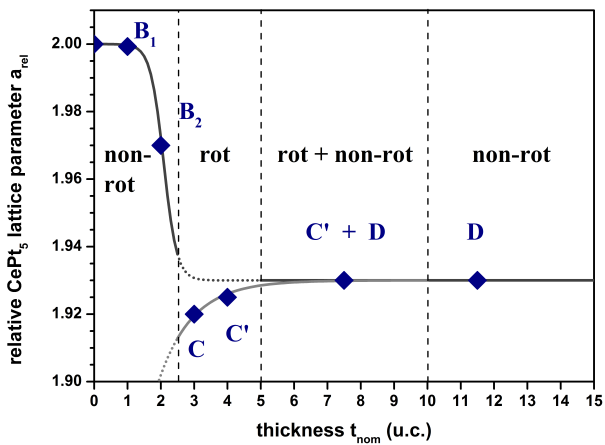

FIG. 4: Evolution of the surface lattice parameter as a function of intermetallic thickness, relative to the one of pristine Pt(111). Symbols represent the lattice parameters of pure phases for $t_{nom} \leq 4$ u.c., and the entire thickness regimes for $t_{nom} > 5$ u.c. Values for $t_{nom} \leq 3$ u.c. are derived from relative LEED spot positions (see text and Ref. 27) the one at $t_{nom} = 4$ u.c. from LEED-IV analysis[42]. Following Ref. 20 we use the CePt5 bulk lattice parameter for $t_{nom} > 5$ u.c.

In the present context, we wish to demonstrate the strong structural similarity of structures $B_2$, C' and D. For this purpose, we display in Fig. 3 the LEED-IV profiles of their first order "$(2 \times 2)$" diffraction spots. The latter were chosen since, independent of thickness, they are unaffected by superimposed substrate diffraction.

The strong similarity between the LEED-IV curves suggests that the building blocks of the underlying atomic structures are essentially the same. Nevertheless, we observe an evolution of higher frequency fine structure with increasing thickness, which we attribute to the fact that a growing number of atomic layers contributes to the diffraction patterns. The important conclusion from our LEED-IV data is that structures $B_2$, C' and D possess the same atomic lattice. Therefore, the variations in electronic structure and magnetic properties reported below occur in one and the same structural phase, probably governed by the variation of the influence of the interface to the Pt(111) substrate.

To summarize the foregoing evaluations, we present the evolution of structure and lattice parameter of the intermetallic film in Fig. 4. It nicely provides an account of stress and strain due to the lattice mismatch between CePt5 and Pt(111) and unites some seemingly contradicting reports concerning lattice relaxation[20,25].

The aligned structure B experiences the tensile stress exerced by the substrate. Its broadened LEED spots hint at the presence of large periodicity superstructures and hence some lattice mismatch which we cannot resolve at $t_{nom} = 1$ u.c. At 2 u.c., the relaxation becomes appreciable and amounts to approx. 1.5%. Upon forma-tion of the rotated structure C at $t_{nom} = 3$ u.c. the lattice parameter reduces to below the CePt5 bulk value, indicative of the compressive stress in this configuration. Further increasing the intermetallic thickness allows lattice relaxation with opposite sign towards the intrinsic CePt5 lattice parameter[20]. In the coexistence regime of structures C' and D we cannot discern a significant difference between their respective lattice constants.

Further insight into film formation and sub-surface atomic structure is obtained from inspecting the lamellar sample extracted from a CePt5/Pt(111) specimen with $t_{nom} = 10$ u.c., shown in Fig. 5. A relatively large intermetallic thickness was chosen to raise the chances of preserving parts of the atomic structure despite the necessity to transfer the specimen under ambient conditions before and after lamella preparation. The LEED pattern (Fig. 5a) exhibits a superposition of the patterns of structures C' and D with lower and higher intensity, respectively corresponding to the high thickness limit of their coexistence. The cross sectional STEM micrograph (Fig. 5b) provides direct evidence for the formation of a film of $\approx 5$ nm thickness, somewhat in excess of the value expected for a 10 u.c. specimen.

In the lower part of the micrograph, a regular pattern of atomic columns of high average brightness is readily recognized. Their arrangement closely matches the pattern expected for an $fcc$ crystal viewed along $\langle 1\,\overline{1}\,0 \rangle$ which is reproduced in Fig. 5d) for reference.

Albeit with lower contrast, atomic structure can still be perceived in parts of the film region, preferentially in the vicinity of the Pt(111) substrate such as in Fig. 5b). The arrangement of atomic columns in the film area reveals a two layer periodicity, compatible with a hexagonal structure. Combining film thickness, periodicity and the amount of Ce deposited leads to the conclusion that the Ce concentration in the intermetallic film amounts to one single atom per $(2 \times 2)$ in every second atomic layer, as expected in case of CePt5 formation. Indeed, associating the visible features with the atomic columns in CePt2 layers on the one hand and with high density columns in the Pt kagome layers on the other, the observed pattern is compatible with the rotated CePt5 structure viewed along the substrate's $\langle 1\,\overline{1}\,0 \rangle$ directions. Across the lamella, this atomic arrangement was found in the vast majority of locations where atomic structure could be resolved.

Analyzing the lateral distances between adjacent atomic columns in more detail, we find their values in the film and substrate, respectively, to coincide to within 1% or less. Since the CePt5 lattice constant closely matches the bulk value, we conclude that the lattice mismatch induced strain is largely accommodated by the Pt substrate in this regime of intermetallic thickness. We recall that for the rotated structure, the stress exerced onto the substrate is of tensile character. Analyzing the evolution of the (1 1 1) atomic layer distance in micrographs covering a larger substrate area than shown in Fig. 5c) we indeed find a continuous increase of the lattice constant with in-creasing distance from the interface, totalling to $\approx 5\%$ over a distance covering 25 (1 1 1) layers and indicative of the relaxation of the tensile interfacial stress.

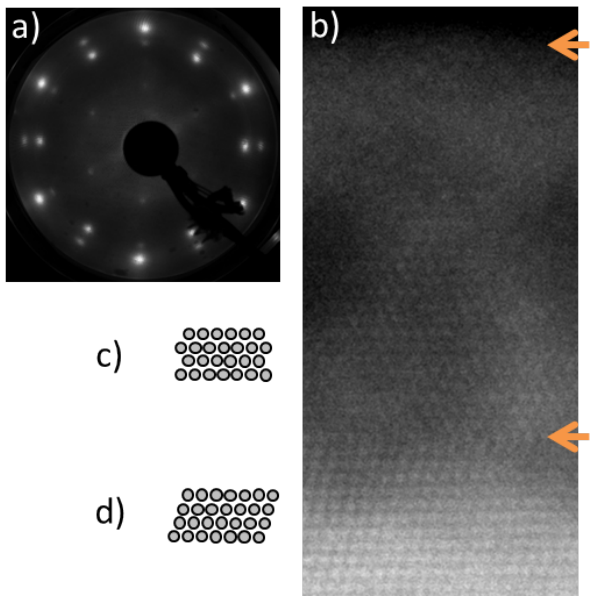

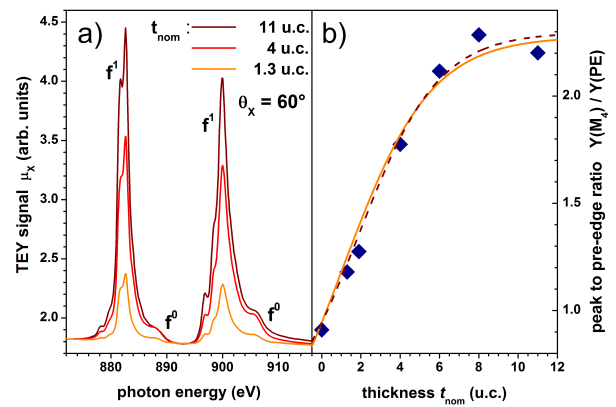

FIG. 5: a) LEED pattern ($E_{kin} = 65$ eV) of a specimen with $t_{nom} = 10$ u.c., taken before extraction of the lamella and showing structure C' as the minority phase and structure D as the majority phase. b) STEM micrograph taken on the lamella of that specimen. Arrows the locations of surface and interface, respectively. The film thickness in the micrograph is about 5 nm, in reasonable agreement with the nominal thickness. c) Symbolic representations of the AB-type stacking of atomic columns found in the film area. d) Expected pattern of an $fcc$ crystal viewed along $\langle 1\,\bar{1}\,0\rangle$.

Owing to the HAADF detection scheme, differences in brightness are primarily related to the density of elements with large atomic number. The ratio of brightness in the film area compared to the substrate amounts to approx. 60%. This value compares favorably to the Pt density ratio of bulk CePt$_5$ vs. the one of pure platinum, which amounts to 64%. In contrast, phases with larger Ce content (e.g. CePt$_{2.23\ldots3}$ as put forward in Ref. 18) possess significantly smaller Pt density. The average HAADF brightness thus provides an independent hint at the film composition which we shall examine with spectroscopic methods next.

Tang et al.[18] have reported Ce-Pt surface alloy compositions determined by XPS, ranging from CePt$_{2.23}$ to CePt$_3$, depending on initial Ce coverage. While this finding actually matches a known homogeneity domain in the bulk phase diagram[28,29,43], it is strongly at variance with the large body of evidence for CePt$_5$ formation at the Pt(111) surface[20,25,27]. Nevertheless, Baddeley et al.[20] report their photoemission results to be in line with those of Ref. 18.

In the present work we use XAS and AES, respectively, to estimate the composition of the surface intermetallic.

FIG. 6: a) CePt$_5$ thickness dependent Ce M$_{4,5}$ XA spectra recorded at x-ray angle of incidence $\theta_X = 60°$ with circular polarization. The high energy shoulders to each of the main multiplets derive from a small admixture of $f^0$ character to the initial state (see text in sect. III B). For clarity, the spectra have been vertically shifted with respect to each other, such that their pre-edge intensities coincide. b) Ratio of M$_4$ resonant (900 eV) to pre-edge (870 eV) intensity as a function of nominal CePt$_5$ thickness (symbols). Solid and dashed lines, respectively, result from model fits (eq. 1), see text for details.

Their respective effective probing depths are essential parameters for the evaluation of near surface specimen stoichiometry. In XAS, this is determined the effective escape depth $\Lambda_T$ of low energy secondary electrons which contribute most of the TEY signal. In AES, the dominant contribution to $\Lambda_A^{eff}$ is given by the inelastic mean free path at the Auger electron kinetic energies instead. Typically, $\Lambda_A^{eff}$ is considerably smaller than $\Lambda_T$[44]. In the absence of quantitative values, $\Lambda_T$ and $\Lambda_A^{eff}$ shall enter as parameters in our models for determining alloy compositions as discussed in more detail below.

Fig. 6a) displays some Ce M$_{4,5}$ XA spectra of specimens with different nominal thicknesses, obtained at temperatures above 250 K and an x-ray angle of incidence of $\theta_X = 60°$, using circular polarized radiation. As discussed before, this choice of parameters ensures that evaluation errors due to linear dichroism are negligible. In line with expectation, the Ce M$_{4,5}$ signal increases monotonically with $t_{nom}$, though non-linearly at elevated thickness.

To obtain an estimate of the composition of the surface intermetallic we proceed by determining the ratio of the TEY signals obtained at the M$_4$ resonance (900 eV) and the pre-edge (PE, 870 eV), respectively (Fig. 6b), and develop a simple model for this ratio as a function of intermetallic thickness. We assume the specimens to be composed of a homogeneous intermetallic film of a priori unknown composition (atomic density $\rho^F$, relative concentrations $c^{Pt}$, $c^{Ce}$ with $c^{Pt} + c^{Ce} = 1$), resting on a semi-infinite Pt substrate. The TEY signal then comprises contributions from both film and substrate, respectively, which depend on film composition and thickness.

The interrelations of atomic absorption cross section $\sigma$ and density $\rho$, absorption coefficient $\mu$ and absorption length $\Lambda_X$ are given by $\Lambda_X^{-1} = \mu = \rho\sigma$. Accordingly, x-rays travelling some distance $t$ in a medium are attenuated by a factor $\Gamma_X(t) = \exp(-t/\Lambda_X)$. To account for the TEY escape depth $\Lambda_T$ we introduce a corresponding attenuation factor $\Gamma_T(t) = \exp(-t/\Lambda_T)$. In these terms, the ratio of TEY intensities at the Ce $M_4$ resonance and the pre-edge, respectively, is given as follows.

$$\frac{I_{M_4}}{I_{PE}} = \frac{\tilde{I}_{M_4}^{Pt}\Gamma_X^{M_4}\Gamma_T + \tilde{I}_{M_4}^F\left(1 - \Gamma_X^{M_4}\Gamma_T\right)}{\tilde{I}_{PE}^{Pt}\Gamma_X^{PE}\Gamma_T + \tilde{I}_{PE}^F\left(1 - \Gamma_X^{PE}\Gamma_T\right)} \qquad (1)$$

where we have abbreviated the unattenuated substrate and film intensities $\tilde{I}^{Pt}$ and $\tilde{I}^F$, respectively, as

$$\tilde{I}^{Pt} \propto \frac{\mu^{Pt}\Lambda_T^{Pt}}{\mu^{Pt}\Lambda_T^{Pt} + \cos\theta_X} \qquad (2)$$

$$\tilde{I}^F \propto \frac{(c^{Pt}\sigma^{Pt} + c^{Ce}\sigma^{Ce})\rho^F\Lambda_T^F}{(c^{Pt}\sigma^{Pt} + c^{Ce}\sigma^{Ce})\rho^F\Lambda_T^F + \cos\theta_X} \qquad (3)$$

Writing TEY intensities as above appropriately takes into account the effect of TEY saturation[45] at x-ray angle of incidence $\theta_X$ with respect to the surface normal. The thickness dependence of the TEY ratio is entirely contained in the attenuation lengths $\Lambda$ and factors $\Gamma$. As indicated, the TEY escape lengths need not be the same in the intermetallic film and the substrate, respectively. The choice of $\Lambda_T^{Pt}$ has little influence on the outcome of our evaluation, however. Guided by the pertinent literature on pure metals[45,46], we have rested with $\Lambda_T^{Pt} \approx 2$ nm. All length scales may readily be expressed in terms of unit cell thickness via $t_{u.c.} \approx 0.44$ nm. Likewise, most other quantities in eq. (1) can be retrieved from the literature and our structural analysis above, reducing it to a model with just $c^{Pt}$ and $\Lambda_T^F$ as free parameters.

Cerium pre-edge and Platinum signals are determined by non-resonant XA absorption coefficients for which we resort to the Henke tables[47] and obtain $\mu_{PE}^{Ce} = 9.9 \ \mu m^{-1}$, $\mu_{PE}^{Pt} = 13.3 \ \mu m^{-1}$, and $\mu_{M_4}^{Pt} = 12.5 \ \mu m^{-1}$ in the pure elements. A quantitative value for the resonant Ce $M_4$ absorption coefficient ($\mu_{M_4}^{Ce} \gtrsim 22.9 \ nm^{-1}$) is extracted from published Ce thin film absorption data measured in transmission[48,49].

Alongside with the experimental data, Fig. 6b) displays fits obtained with eq. (1) which evidently captures the essential behavior of the measurements. In one of the fits (solid line), both $\Lambda_T^F$ and the $c^{Pt}$ were allowed to adopt their optimum values within the least squares fitting procedure, yielding $\lambda_T^F = 0.88(14)$ nm and an alloy composition around CePt$_4$. In a second attempt (dashed line) we fixed the alloy composition to be CePt$_5$ and checked the variation in the peak absorption and probing depth required to reproduce the experimental data.

It turns out that a moderate increase to $\Lambda_T^F = 0.97(32)$ nm along with a 20% rise in peak absorption are required to produce the best fit to the experiment in this case. The relatively large uncertainty in $\Lambda_T^F$ indicates that the influence of this parameter on the outcome of the fitting procedure is limited. Indeed, we shall obtain a somewhat larger value in a related evaluation below. Nevertheless, in accordance with Ref. 6 we emphasize that TEY measurements are fairly surface sensitive, despite contrary statements occasionally appearing in the literature on rare earth containing materials[50,51]. All in all, our XAS results strongly advocate an intermetallic composition with much larger Pt content than previously reported from photoelectron spectroscopy[18].

On the basis of these evaluations, we can also assess the actual importance of TEY saturation for the more extensive evaluations of XAS and XMCD data in the next section. Due to the rather low concentration of Ce in the intermetallic phase, we find that all saturation related sum rule errors are safely contained within 5% at most. We therefore shall not further consider them below.

An analogous set of experiments was performed using AES. For brevity, we restrict the detailed discussion to the regime of large intermetallic film thickness. Fig. 7 displays AES datasets obtained while preparing a specimen with $t_{nom} = 10$ u.c. From bottom to top, the spectra are those of pristine Pt(111), Ce/Pt(111) before and finally after the alloying procedure. As previously noticed[20], the post alloying Auger intensities have converged to the large thickness limit and the substrate no longer contributes to the spectra in this thickness regime. The most prominent Ce and Pt derived low energy Auger features do significantly overlap and are not easily disentangled. We have therefore chosen the Ce contributions at 90/116 eV and the Pt features at 158/168 eV for stoichiometry analysis. All spectra in Fig. 7 have been scaled to equal magnitude of the Pt Auger intensity. It is readily seen that due to the interdiffusion of Ce and Pt upon alloying the relative magnitude of the Ce peaks is considerably reduced. In our analysis we use the published datasets and sensitivity factors of Ref. 52, albeit not without appropriately rescaling the latter according to the strongly different atomic volumes of elemental Ce and Pt[53].

The apparent stoichiometry resulting from straightforwardly applying the sensitivity factors to the topmost spectrum in Fig. 7 yields an apparent composition of CePt$_{10}$. This overestimation of the Pt content essentially results from the Pt termination of the intermetallic, in combination with the small effective Auger electron probing depth $\Lambda_A^{eff}$. The inhomogeneity of the elemental distributions has thus explicitely to be taken into account and stoichiometry determination is not possible independently of a structure model. In our analysis, we therefore rather determine the conditions required to reconcile the apparent composition with the assumption of Platinum terminated CePt$_5$ being formed at the surface.

To this end we construct a model quite analogous to eq. (1), this time explicitely taking the layered atomic structure of CePt$_5$ into account. We formulate the model in terms of atomic layers and accordingly introduce an attenuation factor per layer $\Gamma_{A,L} = \exp\left(-t_L/\Lambda_A^{eff}\right)$, where $t_L = t_{u.c.}/2$ is the distance between adjacent atomic layers along the surface normal in the film. With the relative Ce and Pt densities known in each layer, their respective total Auger intensities are readily expressed as geometrical series. The relative Auger intensities then read

$$\frac{I_{Pt}}{I_{Ce}} = \frac{S_{Pt}}{S_{Ce}} \cdot \frac{1 + 0.5\Gamma_{A,L} - 0.25\Gamma_{A,L}^2}{0.25\Gamma_{A,L}} \qquad (4)$$

where we have incorporated the enhanced Pt density in the terminating layer[34,42]. With the structure model settled, the probing depth $\Lambda_A^{eff}$ thus entirely governs the relation between factual and apparent composition in AES experiments.

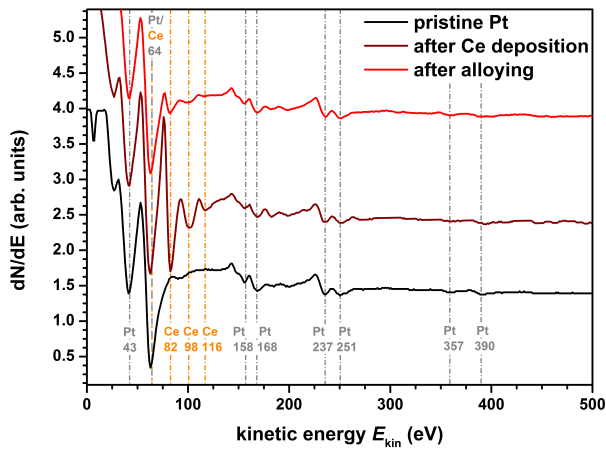

FIG. 7: Sequence of Auger electron spectra acquired with primary energy $E_P = 2.5$ keV during preparation of CePt$_5$ with nominal thicknes $t_{nom} = 10$ u.c. All spectra have been scaled to the same magnitude of the Pt Auger signal around $E_{kin} = 160$ eV and vertically offset for clarity. The strong Ce signatures observed after deposition are strongly reduced after the alloying procedure. Details of stoichiometry evaluation are given in the text.

From eq. (4) we find that the apparent composition CePt$_{10}$ is obtained in case of an effective probing depth of 2.1 atomic layers, i. e. $\Lambda_A^{eff} \approx 4.6$ Å. To obtain the corresponding electron mean free path $\Lambda_{el}$ from this value, we note that $\Lambda_A^{eff}$ does contain additional contributions from the attenuation of the primary electrons ($\Lambda_P \approx 25$ Å) and the take-off angle of the electrons detected in the CMA ($\theta_{CMA} = 42 \pm 3°$). We determine the mean free path from

$$\Lambda_{el}^{-1} = \left(\Lambda_A^{eff\,-1} - \Lambda_P^{-1}\right)\cos\theta_{CMA} \qquad (5)$$

and obtain $\Lambda_{el} \approx 7.6$ Å, which is reasonable number[44].

Modeling the film thickness dependence of relative Ce and Pt Auger intensities in a similar way leads to the same conclusion for the probing depth, i.e. $\Lambda_A^{eff} \approx 7.5$ Å, in order to achieve consistence with CePt$_5$ formation. As with the XA experiments above, our Auger results further corroborate the assumption that CePt$_5$ is the actual composition of the intermetallic compound which forms when Ce is being alloyed into Pt(111).

## B. electronic and magnetic properties

With the structure and composition of the intermetallic phase established, a firm basis is provided for the discussion of its electronic and magnetic properties. As pertinent examples, we report our findings with respect to Ce valence, Kondo temperature and Ce $4f$ paramagnetic response, all of which can be suitably probed by XA experiments.

Due to the finite interaction of the Ce $4f$ states with other electronic states, their occupation shall in general deviate to some extent from $n_{4f} = 1$. Assuming double occupancy to be suppressed by Coulomb repulsion, we may represent the interaction as to result in an admixture of some $f^0$ character to the (predominantly $f^1$) ground state, $|\psi\rangle = c_1 |f^1\rangle + c_0 |f^0\rangle$[54–56]. It is customary to relate the deviation from $n_{4f} = 1$ to the formal Ce valence $V$ via the relation $V = 3 + (1 - n_{4f})$. As in photoemission, its presence manifests itself through the appearance of additional spectral signatures in XA spectra[6,55–59], whose intensities vary proportionally to the magnitude of the $f^0$ admixture to first order. In the Ce $M_{4,5}$ XA spectra of Fig. 6 these are visible as shoulders on the high energy side to both resonances.

Tang et al.[18] already noticed a non-monotonous dependence of the Ce $4f$ valence on Ce dose in the surface alloyed Ce-Pt phases. In the present work, we proceed to a similar analysis on the basis of low temperature ($T \lesssim 20$ K) XA spectra, otherwise taken under the same experimental conditions as the datasets of Fig. 6. Low temperature spectra were chosen to capture the ground state properties to the best possible extent. The evaluation procedure consists of determining the relative spectral weight $w_{rel}^{0\rightarrow1}$ of the $f^0$ contribution to the XA spectra. For convenience, we define the latter as relative to the $f^1$ spectral weight, i.e. $w_{rel}^{0\rightarrow1} = w^{0\rightarrow1}/w^{1\rightarrow2}$. Numerical simulations of the $f^1$ absorption multiplet using the 'Solid State Physics Mathematica Package'[60] (not shown) were used to aid the quantitative extraction of $w_{rel}^{0\rightarrow1}$. Neglecting many-body corrections[56], $n_{4f}$ is then given as

$$n_{4f} = (1 + w_{rel}^{0\rightarrow1})^{-1} \qquad (6)$$

Fig. 8 displays the result of these evaluations along with those of Tang et al.[18], for which we have appropriately[27] converted the reported Ce depositions into thickness values of the resulting surface intermetallic. The good agreement between the two datasets is obvious and surpasses theoretical expectations, which in principle yield different proportionality factors between ground state admixture and spectral weights for XPS and XAS, respectively[55,56].

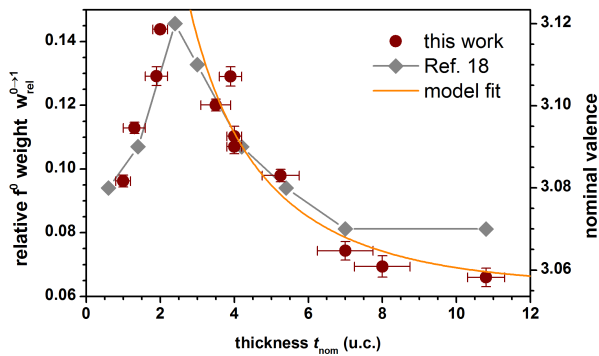

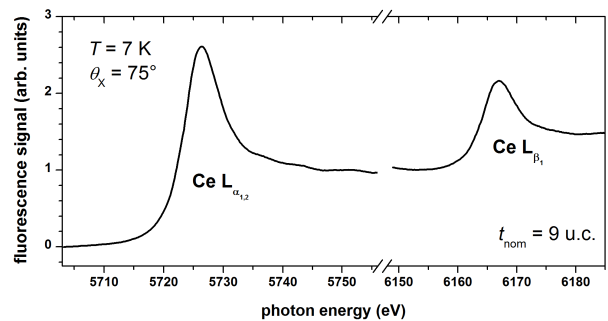

FIG. 8: Thickness dependence of relative spectral weight $w_{rel}^{0 \to 1}$ (left scale) and corresponding nominal valence (right scale). Along with our evaluation of XAS data (circles) we included the photoemission results of Tang et al., extracted from Ref. 18 (diamonds), to which only the right scale applies. The good agreement between the two datasets is noticeable. The solid line appearing for $t_{nom} > 3$ u.c. represents the model fit according to eq. (7), as discussed in the text.

FIG. 9: Low temperature Ce $L_{2,3}$ absorption data of a CePt$_5$/Pt(111) specimen with $t_{nom} = 9$ u.c., capped by 4 ML of Pt. Note the absence of spectral features at energies above the white lines, which would derive either from oxidation or a significant $f^0$ admixture to the ground state, as observed in mixed valence compounds.

It is noticeable that the peak in valence nearly coincides with the transition from the non-rotated to the rotated CePt$_5$ phase. We therefore reckon that different mechanisms are responsible for the increasing valence at small intermetallic thickness and its subsequent decrease at larger thickness, respectively. The initial behaviour is reminiscent of the pressure induced, isostructural $\alpha$-$\gamma$ transition of elemental Ce, where the increase in valence goes along with a substantial reduction of the lattice parameter. Since we observe a sizeable reduction of the lattice parameter of the CePt$_5$ surface intermetallic upon increasing its nominal thickness from 1 u.c. to 2 u.c., we conclude by analogy that the concommitant increase of valence represents just another aspect of this structural relaxation.

Above 4 u.c. the CePt$_5$ lattice parameter undergoes only very little change. The observed decrease of Ce valence could instead arise as the weighted average of different local values, governed, e.g., by the proximity to the film surface or the interface to the Pt(111) substrate, where the Ce atoms possess an increased number of Pt neighbors with respect to the CePt$_5$ lattice. A similar scenario was proposed by Rothman et al. in their analysis of epitaxial Ce films[61]. To test the validity of our hypothesis we again set up a simple model equation for the measured relative $f^0$ spectral weight $w_{rel}^{0 \to 1}$. Very much in analogy to the preceding cases, the probing depth will play an essential role and determine the spectral contributions from different atomic layers. In our model we assume that the total spectral weight $w_{tot} = w^{1 \to 2} + w^{0 \to 1}$ is independent of Ce valence, a simplification which appears justified in view of the small changes in $n_{4f}$ incurred. Our expression for $w_{rel}^{0 \to 1}$ reads as follows:

$$w_{rel}^{0 \to 1} = \frac{w_s^{0 \to 1}\Gamma_s + w_b^{0 \to 1}\Gamma_b + w_i^{0 \to 1}\Gamma_i}{w_s^{1 \to 2}\Gamma_s + w_b^{1 \to 2}\Gamma_b + w_i^{1 \to 2}\Gamma_i} \qquad (7)$$

where the indices classify the Ce sites within each atomic layer as being either surface or interface determined (s and i, respectively) or unaffected by both ('bulk-like', b). The $\Gamma$ factors describe the relative weight of the respective contributions due to the TEY escape length and relate to the thicknesses $t_s$, $t_b$ and $t_i$ of the three different zones according to $\Gamma_s = 1 - \exp(-t_s/\Lambda_T)$, $\Gamma_b = \exp(-t_s/\Lambda_T) - \exp(-t_b/\Lambda_T)$ and $\Gamma_i = \exp(-t_b/\Lambda_T) - \exp(-t_i/\Lambda_T)$. Evidently, the total intermetallic thickness $t$ must be given by $t = t_s + t_b + t_i$.

We have represented in Fig. 8 the simplest version of eq. (7) yielding a satisfactory fit to the experimental data for $t \gtrsim 4$ u.c. as a solid curve. It is characterized by attributing single unit cell thickness to both surface and interface regions, respectively, which additionally share the same value for the valence, hence $w_s^{0 \to 1} = w_i^{0 \to 1}$. The optimization yields $w_s^{0 \to 1} = w_i^{0 \to 1} = 0.17$, $w_b^{0 \to 1} = 0.021$ and an effective probing depth of $\Lambda_T = 3.4$ u.c. (1.5 nm). This value is larger than the one we obtained from fitting eq. (1) to the data in Fig. 6 above. However, considering the error bars and the simplicity of both model equations, the discrepancy is acceptable and does not significantly affect our conclusions. In particular, we see that the consideration of surface and interface effects provides a plausible scenario to account for the observed trend in the Ce valence and highlights the fact that it is imperative to account for the finite probing depth in evaluations of TEY data. Moreover, while our model is certainly crude, the smallness of $w_b^{0 \to 1}$ may well provide a clue as to why bulk CePt$_5$ was previously classified as a material with negligible Kondo screening[31].

This assumption is supported by the low temperature Ce $L_{2,3}$ absorption spectrum displayed in Fig. 9, acquired on a (Pt capped) specimen with $t_{nom} = 9$ u.c. Due to both, the much larger x-ray penetration length and fluo-rescence detection scheme, the hard x-ray spectrum represents a homogeneous probe of the intermetallic film. As in Ce $M_{4,5}$ data, a significant admixture of $f^0$ character to the ground state would lead to the appearance of a distinct spectral signature at higher excitation energy[62]. As is evident from Fig. 9, such a feature cannot be discerned in our data, indicating that the average $f^0$ weight is small at this film thickness.

The presence of (impurity) Kondo scattering can be probed by XAS by monitoring the $f^0$ related spectral weight $w_{rel}^{0 \to 1}$ as a function of temperature. Numerical solutions to the single impurity Anderson model (SIAM) predict an increase of $n_{4f}$ as the temperature is raised from (well) below the Kondo temperature to (well) above[63,64]. The associated change in $w_{rel}^{0 \to 1}$ proves way inferior to its thickness dependent variation in our specimens discussed above. Fig. 10 gives two examples of our observations, displaying $w_{rel}^0(T)$ and associated (average) $\Delta n_{4f}$ for two specimens with a thickness of 1 u.c. and 4 u.c., respectively[65]. These were chosen for their comparable $w_{rel}^{0 \to 1}$ and hence valence at low temperature (see Fig. 8).

Nevertheless, both specimens feature distinct characteristic temperature scales in their increase of $w_{rel}^{0 \to 1}$, revealing some difference in the details of the interaction between $4f$ and itinerant electrons. As a rough estimate of the Kondo scale, we associate it with 50% of the change in $n_{4f}$ between the lowest and highest temperatures covered by our experiment. This approximately coincides with the region of largest average slope in $n_{4f}(T)$. In this way, we obtain $T_K \approx 125$ K at 1 u.c. and $T_K \approx 200$ K at 4 u.c., respectively. The strong difference between the slopes of $\Delta n_{4f}$ for 1 u.c and 4 u.c. is largely due to the respective values of $T_K$, resulting in very similar $\Delta n_{4f}$ vs. $T/T_K$ behavior. Yet, the larger scatter of the experimental data at $t_{nom} = 1$ u.c. readily creates the impression of an untypically abrupt increase in $\Delta n_{4f}$. We note in particular that $T_K \approx 200$ K at 4 u.c. readily accounts for the observability of Kondo resonance features at $T \approx 120$ K in photoemission[19].

Finally, we shall discuss the magnetic response of our CePt$_5$ films as observed with XMCD. The Ce $4f$ local moment paramagnetic response is largely dependent on energetic order and splitting of the low-lying Ce $4f$ states with $j = 5/2$. In a hexagonal crystal field (CF), these states generally split into three Kramers doublets, given by the $|\pm m_j\rangle$ states with $m_j = 1/2$, $3/2$ and $5/2$[66]. In bulk CePt$_5$, the $m_j = 1/2$ Kramers doublet has been determined to be the ground state, the first CF excited state lying $\Delta_{CF} = 27$ meV higher in energy[35]. For $k_B T \ll \Delta_{CF}$ the unscreened Ce $4f$ paramagnetic moment along the hexagonal axis then amounts to $m_{eff} = g\sqrt{3/4}\mu_B$ with Landé factor $g = 6/7$.

In the presence of Kondo scattering, $T_K$ sets the scales of both the change of impurity occupation and the screening of its effective paramagnetic moment[64]. As a result, the magnetic response gradually changes from an asymptotically free moment Curie-Weiss-like (CW) be-havior for $T \gg T_K$ towards a finite, Pauli-like susceptibility as $T \to 0$. This generic behavior is well accounted for by the single impurity Anderson and Kondo models and remains essentially valid at elevated impurity density as long as $T \gtrsim T_K$. In Kondo lattices, the coherence temperature $T^*$ may emerge as yet another temperature scale[67]. It characterizes the onset of interactions between the $4f$ sites which ultimately lead to the formation of heavy fermion bands as $T \ll T^{*}$[68].

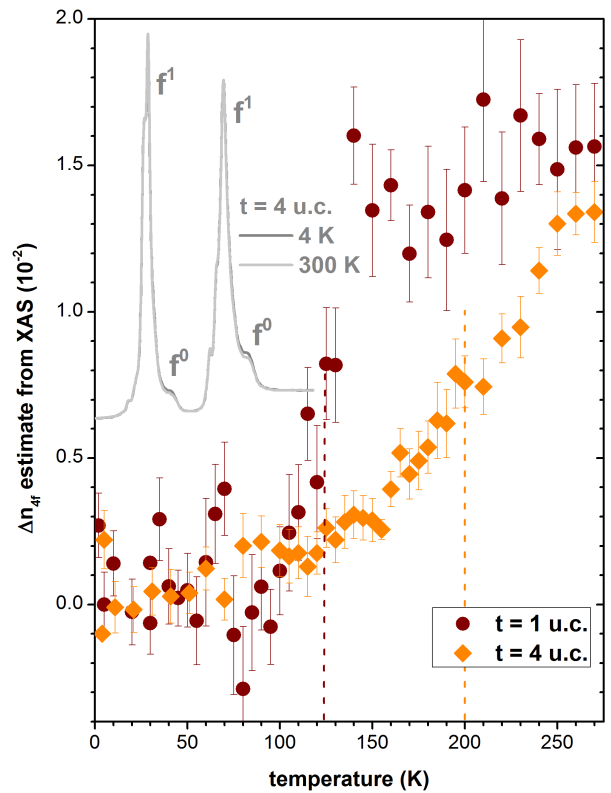

FIG. 10: Temperature dependence of the Ce $4f$ occupation numbers, estimated from the corresponding changes of $w_{rel}^{0 \to 1}$ while raising the specimen temperature from below 10 K to 275 K. CePt$_5$ film thicknesses of 1 u.c. and 4 u.c. were chosen due to their similar effective valence at low temperature (Fig. 8). The inset displays (without scales) the low and high Ce $M_{4,5}$ spectra for $t_{nom} = 4$ u.c., scaled to same intensity at the $M_4$ resonance, highlighting the variation in $w_{rel}^{0 \to 1}$.

Since $T^* \lesssim 20$ K in CePt$_5$/Pt(111) at 4 u.c. thickness according to Ref. 26, we identify the temperature range $T^* \lesssim T < T_K, T_{CF}$ as a most interesting one, where we may expect the Ce $4f$ moments to display a Curie-Weiss-like susceptibility with screened moments. We therefore focus on this range in what follows and use the XMCD at the Ce $M_{4,5}$ edges to determine the Ce $4f$ susceptibility and effective paramagnetic moments with the magnetic field aligned along the surface normal[69]. Fig. 11a) displays a Ce $M_4$ XA spectrum of a 4 u.c. specimen along with dichroism datasets measured at various temperatures. As expected for a local moment paramagnet, the XMCD signal is strongest at lowest temperature and rapidly decreases towards higher temperature.

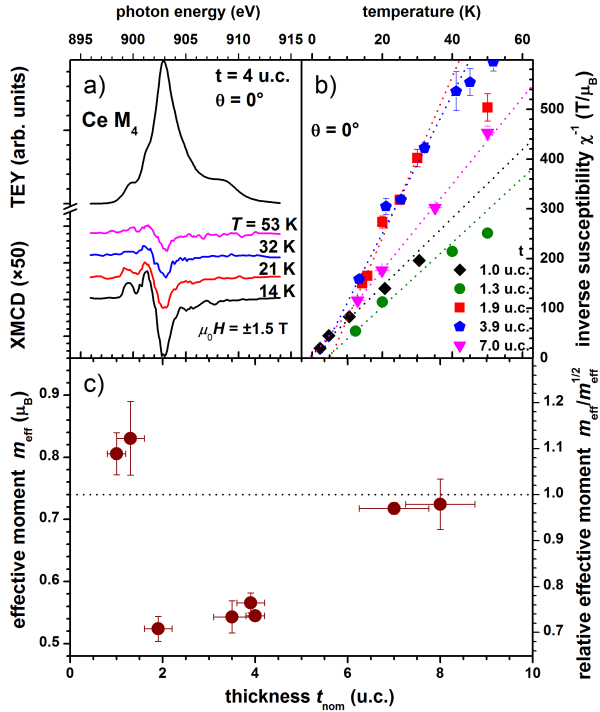

FIG. 11: a) Ce $M_4$ XA and XMCD datasets of CePt$_5$ at 4 u.c. thickness. The XA spectrum is representative for all temperatures. XMCD data for a number of temperatures are given on an expanded scale and have been vertically offset for clarity. b) Inverse Ce $4f$ susceptibility derived from XMCD datasets such as in panel a). Dashed lines indicate respective Curie-Weiss laws derived from fitting the inverse susceptibility data. c) Effective Ce $4f$ moments as a function of CePt$_5$ thickness, derived from the Curie-Weiss fits. The horizontal line indicates the effective Ce $4f$ moment expected for $m_j = 1/2$ in the absence of Kondo screening. Values fall well below this line in the region of enhanced valence and indicate the presence of Kondo screening. Enhanced moments at small CePt$_5$ thickness are indicative of small crystal field splitting, leading to finite contributions of a higher lying level with $m_j > 1/2$.

Fig. 11b) displays the inverse Ce $4f$ susceptibilities of a selection of specimens, obtained from the XMCD data as outlined in the appendix. Very obviously, the CePt$_5$ specimens show quite substantial variations in their Ce $4f$ response. For each specimen we approximate the low temperature data ($T \lesssim 40$ K) by a Curie Weiss law to obtain effective magnetic moments. Towards higher temperature, deviations from CW behavior is observed, due to non-negligible thermal population of CF excited states. The moments $m_{\text{eff}}$ follow from the susceptibility via the usual relation

$$\chi^{-1}(T) = \frac{3k_{\text{B}}(T - \Theta_{\text{p}})}{m_{\text{eff}}^2} \qquad (8)$$

where $\Theta_{\text{p}}$ is the paramagnetic CW temperature. We in-variably find $\Theta_{\text{p}} > 0$ which is an indication of predominantly ferromagnetic correlations, presumably mediated by RKKY interaction.

The resulting Ce $4f$ effective moments are plotted in Fig. 11c) as a function of CePt$_5$ thickness. The plot reveals a remarkable thickness dependence. Towards large thickness $m_{\text{eff}}$ approaches the expected value of $m_{\text{eff}} = 0.74\mu_B$. This value is also represented by a horizontal dotted line in the plot.

By contrast, the effective Ce $4f$ moments are markedly reduced in the range of $2 \ldots 4$ u.c. In principle, such a reduction could arise from strong antiferromagnetic correlations of at least a subset of the Ce atoms in the specimens. We lack a conceivable mechanism, however, to produce such strong Ce-Ce coupling and as noted above, the dominant correlations rather appear to be of ferromagnetic character. We therefore conclude that the reduced $m_{\text{eff}}$ provide direct evidence for Kondo screening in the CePt$_5$ surface intermetallic.

On a more quantitative level, the screening we observe amounts approximately to the strength expected around $T \approx T_{\text{K}}$ in the single impurity limit[1,63,64]. Since we determined the effective moment at considerably lower temperature, it follows that the Kondo screening in the CePt$_5$ intermetallic is weaker than for single impurities – a fact which, since the relevant conduction band significantly departs from half filling[26], could bear some relationship to Nozière's famous exhaustion problem[70]. The experimental observation of $T^* \ll T_{\text{K}}$ would actually fit to such a scenario[67,71].

Towards smallest CePt$_5$ thickness, we find a resurgence of $m_{\text{eff}}$ to even above the unscreened value for $m_j = \pm 1/2$. Such increased moments can only be accounted for by assuming a non-zero contribution of higher $m_j$ states to the Ce $4f$ susceptibility and thus the occurrence of strongly reduced CF splitting compared to bulk CePt$_5$. Since these moments arise in the thickness range of largest strain, such a modification of the CF level scheme is readily conceivable. Nevertheless, the relative smallness of the excess moment suggests that either the corresponding Boltzmann weight $\exp(-\Delta_{\text{CF}}/k_B T)$ is relatively small or that Kondo screening is sizeable in this Kramers doublett just as well (or both). Given its finite Kondo temperature, the single CePt$_5$/Pt(111) layer in any case represents an experimental realization of a two-dimensional Kondo lattice deserving further study by means of surface science techniques.

## IV. SUMMARY

In conclusion, we have presented a comprehensive study of film formation and composition as well as electronic and magnetic properties of the Ce-Pt surface intermetallics forming upon alloying Ce into Pt(111). We have provided evidence for the fact that well-defined intermetallic films form over a thickness range up to (and above) 5 nm. Irrespective of the variety of LEED pat-terns as a function of Ce dose, the atomic structure may throughout be understood to consist of the $CePt_5$ lattice with thickness dependent distortions due to interfacial stress. Scattering intensity in STEM as well as compositional analysis by AES and XAS are in accordance with the assumption of $CePt_5$ formation. $CePt_5/Pt(111)$ is therefore appropriately be referred to as a surface intermetallic compound.

The structural variations as a function of intermetallic thickness are accompanied by systematic changes in the electronic and magnetic properties, which are readily captured by soft x-ray absorption experiments. Our analysis suggests that the increase of Ce valence at low $t_{nom}$ is primarily driven by the relaxation of the lattice parameter, while its subsequent decrease at larger thickness rather reflects a dependence of Ce hybridization on the local number of Pt neighbors.

Monitoring the Ce valence as a function of temperature by XAS reveals the presence of Kondo scattering and allows a rough estimate of the Kondo scale. It amounts to $T_K \approx 125$ K at $t_{nom} = 1$ u.c. and $T_K \approx 200$ K at $t_{nom} = 4$ u.c. By means of XMCD at the Ce $M_{4,5}$ edges, we are able to specifically address the temperature dependent paramagnetic Ce $4f$ response in a temperature range between the lower and higher energy scales set by the coherence temperature ($T^* \lesssim 20$ K) on the one hand, and by the impurity Kondo temperature ($T_K \propto 10^2$ K) and crystal field excitations on the other. In the thickness range of $t_{nom} = 2\ldots4$ u.c. in particular, these measurements allow a direct observation of the partially screened Ce $4f$ moments and reveal the magnitude of screening in these Kondo lattices to lag behind simple theoretical expectations for single impurity screening. We attribute the resurgence of excess moments at smallest $CePt_5$ thickness to a strong modification of the crystal field splittings under the constraint of large tensile lattice strain.

Nevertheless, $CePt_5/Pt(111)$ proves to be a structurally robust Kondo lattice system with a remarkable electronic tunability, readily implemented by varying the thickness of this surface intermetallic. The XMCD technique provides a direct and sensitive magnetic probe of the local sites and the interactions with their environment. It therefore constitutes a most useful complementary method for studying strongly correlated electron systems at surfaces.

## Acknowledgments

It is a pleasure to acknowledge M. Haverkort and P. Hansmann for their valuable input and support concerning simulations of XA spectra. We thank F. F. Assaad and M. Bercx, F. Reinert and H. Schwab, as well as G. Held for helpful discussions. This work was financially supported by the Deutsche Forschungsgemeinschaft within research unit FOR1162 (P7). Access to synchrotron radiation was also partially funded by BMBF (contract no. 05ES3XBA/5) and the HZB (BESSY II), and by the European Community's Seventh Framework Programme (FP7/2007-2013) under CALIPSO project (grant agreement no. 226716). The ESRF beamtime was granted via proposal HE-3443. Generous allocation of beamtime at synchrotron radiation facilities as well as their general support is gratefully acknowledged.

## Appendix: inverse XMCD susceptibility

The temperature dependent (inverse) susceptibility was usually determined as follows (datasets taken at BESSY II). For each specimen, a full Ce $M_{4,5}$ XMCD dataset was measured at the lowest temperature available in the respective experiment. The Ce $4f$ orbital magnetic polarization was determined by applying the usual orbital moment sum rule[72–75]. The spin moment sum rule being unlikely to be applicable in Ce[76–78], we have instead calculated it using the atomic Landé factor and thus assuming the relation $m_S = -m_L/4$ between spin and orbital moments to hold. Since the applied field ($\mu_0 H = \pm 1.5$ T) lies in the linear response regime of our specimens, the atomic susceptibility (in $\mu_B$ atom$^{-1}$ T$^{-1}$) can directly be inferred from this evaluation. The respective values at more elevated temperatures were obtained by comparing the relative magnitudes of the Ce $M_4$ XMCD to the lowest temperature one. The latter procedure relies on the assumption that the dichroic line shape does not change with temperature, which was thoroughly checked in quite a number of cases.

The procedure adopted for data taken at SOLEIL was different due to the fact that XMCD spectra were acquired at larger applied fields ($\mu_0 H = \pm 4\ldots6$ T), where the magnetic response is nonlinear at the lowest temperatures. XMCD magnetization curves were then additionally acquired to derive the zero field susceptibility.